# Nanoemulsions obtained via bubble bursting at a compound interface

Jie Feng[1], Matthieu Roché[1, #], Daniele Vigolo[1, $], Luben N. Arnaudov[2], Simeon D. Stoyanov[2,3],

Theodor D. Gurkov[4], Gichka G. Tsutsumanova[5] and Howard A. Stone[1, *]

1. Department of Mechanical and Aerospace Engineering, Princeton University, Princeton, New Jersey, 08544, USA
2. Unilever Research and Development, 3133AT Vlaardingen, The Netherlands
3. Laboratory of Physical Chemistry and Colloid Science, Wageningen University, 6703 HB Wageningen, The Netherlands; Department of Mechanical Engineering, University College London, Torrington Place, London WC1E 7JE, UK
4. Department of Chemical Engineering, Faculty of Chemistry & Pharmacy, University of Sofia, Sofia 1164, Bulgaria
5. Department of Solid State Physics & Microelectronics, Faculty of Physics, University of Sofia, Sofia 1164, Bulgaria

---

[#] Present address: Laboratoire de Physique des Solides, Université Paris Sud-CNRS, 91405 Orsay, France

[$] Present address: Department of Chemistry and Applied Biosciences, Institute for Chemical and Bioengineering, ETH Zurich, 8093 Zurich, Switzerland

[*] Corresponding author: hastone@princeton.edu

**The bursting of bubbles at an air/liquid interface is a familiar occurrence important to foam stability[1], cell cultures in bioreactors[2] and mass transfer between the sea and atmosphere[3-4]. Here we document the hitherto unreported formation and dispersal into the water column of submicrometre oil droplets following bubble bursting at a compound air/oil/water-with-surfactant interface. We show that dispersal results from the detachment of an oil spray from the bottom of the bubble towards water during bubble collapse. We provide evidence that droplet size is selected by physicochemical interactions between oil molecules and the surfactants rather than by hydrodynamic effects. We illustrate the unrecognized role that this dispersal mechanism may play in the fate of the sea surface micro-layer and of pollutant spills by dispersing petroleum in the water column. Finally, our system provides an energy-efficient route, with potential upscalability and wide applicability, for applications in drug delivery[5], food production[6] and material science[7], which we demonstrate by producing polymeric nanoparticles.**

Previous studies of bubbles bursting at an air/water interface investigated mass transfer from a lower liquid phase to a upper gas phase[8-10], which also occurs when a rising bubble passes through an oil/water interface[11]. Here, we describe the reverse transport process, where submicrometre oil droplets, formed during bubble bursting at a compound interface are transported from the upper to the lower phase. We are not aware of any previous documentation of this phenomenon. After continuous bubble bursting at an air/hexadecane/water-with-surfactant interface for dozens of hours (Fig. 1a), the aqueous phase became translucent, which suggested that small objects had been dispersed in the lower water phase (Fig. 1b). The analysis of samples of the solution using dynamic light scattering (DLS) confirmed the presence of objects with a radius $r \sim$ 100 nm and a moderate polydispersity index (PDI, see Supplementary Materials) (Fig. 1c). Since the surfactant concentration in the water phase is well below the critical micelle concentration[12] and we only observe submicrometre objects when there is bubble bursting, these objects are oil droplets rather than spontaneously generated microemulsion droplets or surfactant mesophases. Control experiments also confirm that these submicrometre-sized droplets exist only when surfactants are present in the water phase. Measurements of the size of hexadecane droplets on longer timescales showed that $r$ remained constant for at least a week (Fig. 1d). Thus, our experiments demonstrate that the bursting of air bubbles at a compound interface also drives mass transport into the bulk liquid to form stable submicrometre droplets.

High-speed visualization of the bubble bursting process from above the air/oil interface and below the oil/water interface allowed us to understand how oil droplets are dispersed in the surfactant solution. The bubble cap, initially formed of an oil film sitting on a water film, bursts in two steps in most cases (Fig. 2a, Supplementary Movie 1). The experimental images suggest that the top oil film retracts first and then the water film breaks. Only the latter step induces droplet production through fragmentation of the receding film, thus ruling out atomization of the bubble cap as the origin of the submicrometre-sized oil droplets in the aqueous phase. A side view of the bursting bubble below the oil/water interface reveals that oil droplets are dispersed from the bottom of the bubble into the bulk water (Supplementary Movie 2). We have observed that after a hole opens in the water film (blue circle in Fig. 2b.1) the surface of the cavity deforms during film retraction (Fig. 2b.2) and a spray of droplets is ejected from the cavity boundary towards the bulk water, at a location opposite to that of the nucleation site of the hole (red circle in Figs. 2b.3-4). Also, we observed that the larger non-Brownian droplets rose rapidly back to the oil/water interface, while the smaller objects were ejected deep in the bulk water (Supplementary Movie 3). We speculate that submicrometre-sized droplets are formed and dispersed during this spraying process, although we cannot observe them directly with optical methods.

During the time that we observe droplet formation, the flow close to the bubble-water interface resembles the boundary-layer detachment flow theoretically predicted for the case of a bubble bursting at an air/water interface[13]: the separation of the hydrodynamic boundary layer around the cavity means that streamlines detach nearly perpendicular to the bubble boundary, which in our case leads to the spray into the bulk fluid. We performed additional model experiments that show that 10-µm latex particles initially sitting on the flat air/water interface without the oil phase are also dispersed from the side of the cavity into the bulk water during bubble bursting in a fashion similar to the oil droplets (Supplementary Movie 4). Thus, we propose the dispersal mechanism summarized in Fig. 2c. For a bubble at the compound interface, the upper oil film ruptures first, leaving a water film that retracts rapidly after a hole nucleates on the bubble cap (Figs. 2c.1-3). Then, a spray of polydisperse droplets is generated (Fig. 2c.4).

The similarity of the dispersal mechanism with the predicted boundary-layer detachment motivated us to investigate how hydrodynamics may set the size of submicrometre droplets. However, a study of this relationship with more than ten non-aqueous phases and three surfactants in the aqueous phase (see Methods) shows that the results are different from most fragmentation processes[14-15]. In particular, since the size of the droplets $r$ we measured is independent of a change of the initial thickness of the oil layer $h_I$ or the bubble diameter $d_b$ (Figs. 3a,b), we used dimensional analysis to determine a characteristic length scale for our system, which depends on viscosity of the oil phase $\eta_o$, viscosity of the aqueous phase $\eta_w$, density of the oil phase $\rho_o$ ($\approx \rho_w$) and the interfacial tension between the oil and water $\gamma_{ow}$. Since $\eta_o \geq \eta_w$, we assume that only $\eta_o$ is significant, and we obtain $r \sim \eta_o^2/(\rho_o \gamma_{ow})$. Unfortunately, this naive scaling law fails to capture the three to five-fold decrease of $r$ with a three-fold increase of $\eta_o$ (Fig. 3c). In addition, we observed that $r$ increases with an increase of the speed of the last receding film $U_r$ (Fig. 3d), in contradiction with the expectation that an increase of the energy injected in a two-phase liquid system generates smaller droplets[15]. Thus, the size of the submicrometre droplets is independent of hydrodynamics.

We hypothesize that the submicrometre droplet size is set by molecular-scale physicochemical interactions between the oil molecules and the surfactants. At the macroscopic scale, these interactions translate into transitions between three possible

102 wetting states for oil on an aqueous surfactant solution depending on surfactant

103 concentration: partial, pseudo-partial and complete wetting[16-18] (Fig. 4a). We found that

104 dispersal of submicrometre-sized droplets never occurred for liquid combinations

105 showing only a pseudo-partial-to-complete wetting transition (poly(dimethylsiloxane)

106 on aqueous surfactant solutions[19]) or partial wetting for all surfactant concentrations

107 (alkanes on aqueous solutions of Aerosol OT[16]). In contrast, dispersal occurred in

108 systems where only a surfactant-induced transition from partial to pseudo-partial wetting

109 happened. We observed the presence of oil droplets either when the equilibrium

110 surfactant concentration in water was high enough to induce a pseudo-partial wetting

111 state at rest, or when the surfactant concentration was smaller than the transition

112 concentration, but sufficiently close to it, so that surfactant compression[20] during bubble

113 bursting dynamics could trigger a wetting transition.

114 The correlation between the oil/water wetting state and the occurrence of dispersal

115 also suggests a possible explanation for the origin of the droplets. In our study, systems

116 showing pseudo-partial wetting involve linear alkanes and surfactants. Linear alkanes,

117 which are structurally similar to the hydrophobic moiety of the surfactants, can penetrate

118 the surfactant monolayer, and shorter alkanes penetrate more readily than longer alkanes.

119 As a consequence, the size of submicrometre alkane droplets obtained using a given type

120 of surfactant would increase as the length of the alkane carbon chain decreases[9,21],

121 similar to what we observed (Fig. 3c). In addition, pseudo-partial wetting is

122 characterized by the coexistence of oil lenses at equilibrium with a thin film of oil whose

123 thickness is on the order of several molecular sizes[17]. These lenses could be the seeds of

124 our submicrometre droplets. To test this idea, we performed ellipsometry measurements

125 after the deposition of a millimeter-sized hexadecane droplet on a surfactant solution,

126 which show that small patches with a thickness on the order of 100 nm exist at places on

127 the interface (see Methods).

128 Next, we use this experimental result to deduce the lateral size and hence the volume

129 of the lenses so as to compare with the volume and the radius of the

130 submicrometre-sized droplets. The wetting state of oil/water+surfactant systems results

131 from a competition between short-range and long-range interactions described by the

132 initial spreading coefficient $S_i$ and the Hamaker constant $A$, respectively[17]. Balancing

133 long-range van der Waals interactions with surface tension in a small slope

134 approximation gives

$$A/h^3 \approx S_i h/\lambda^2, \qquad (1)$$

136 where $h$ is the lens thickness, which is on the order of 100 nm, and $\lambda$ is the lateral size of

137 the lens (Fig. 4a; Supplementary Materials). Then, $\lambda \approx O(10~\mu m)$, which is consistent

138 with other studies in pseudo-partial wetting[22]. The volume of one single lens is $V_{lens} \approx$

139 $h\lambda^2$ and then we expect the droplet volume $r^3 \approx h\lambda^2$. Hence $r \approx (h\lambda^2)^{1/3} \approx 10^{-6}$ m, which is

140 the order of magnitude of the size we measure. Direct confirmation of our hypotheses is

141 difficult experimentally due to the millisecond time scales and the submicrometre length

142 scales characteristic in our system, but the above arguments are consistent with all of our

143 observations.

144  The results we report here have important environmental consequences. For
145 example, we have verified that petroleum is dispersed in bulk water by bubble bursting
146 (Figs. 4b,c). For those environments where other well-identified mechanisms are limited,
147 the bubble-bursting process is one possibility to disperse oil into the lower water phase.
148 This dispersal may enhance pollution, where small droplets tend to be digested by sea
149 creatures more easily than on the surface, but dispersal may also help bacteria or algae to
150 degrade pollutants faster[23] because of the high surface-to-volume ratio of the droplets.
151 Also, the structural similarity between the compound interface that we study and the
152 interface separating the ocean from Earth's atmosphere, which is always covered by the
153 sea surface micro-layer (SML) containing surface-active organic matter[24], suggests that
154 the SML can be transported into the bulk of the oceans by bursting bubbles[25]. For the
155 above scenarios, we are not aware of any study investigating mass transfer from the
156 surface of the ocean towards its bulk due to bubble bursting, which has been related so
157 far only to the formation of wind-dispersed aerosols [26].

158  Inspired by the application of bubbling to the production of colloids[27] and
159 liposomes[28], our study provides a potential scalable route for the production of
160 dispersions of submicrometre particles. As an illustration, we have dispersed droplets of
161 a polymer liquid (NOA 89) and we cross-linked it using UV light to obtain solid
162 particles with a size comparable to that of the original droplets (Figs. 4d,e; Methods).
163 Our dispersal method meets three requirements important to industry. First, its energy
164 efficiency is 1 - 10%, which is at least one order of magnitude greater than the $O(0.1\%)$

165 efficiency of classical high-shear-rate methods[29] (see Supplementary Materials for
166 detailed calculations). Second, bubble bursting has potential to be scaled up, by
167 increasing bubbling frequency for example (see Supplementary Materials), while
168 keeping costs low and remaining sustainable, in contrast with the mechanical top-down
169 methods. Finally, compared to classical self-emulsification for nanoemulsions[30], which
170 only works for specific compounds having ultra-low interfacial tensions, our system
171 works even when interfacial tensions are on the order of tens of mN/m and thus it has
172 broad applicability.
173
174
175
176
177
178
179
180
181
182
183
184
185

## Methods

**Experimental system** The experimental system is shown in Fig. 1a. For each experiment, a thin layer of the dispersed phase, e.g. a non-polar oil, was deposited on an aqueous solution containing an ionic surfactant $C_n$TAB. Air bubbles were formed at the tip of a tube located at the bottom of the tank. The bubbles rose to the interface because of buoyancy. We changed the size of the bubbles by adjusting the injection pressure and the diameter of the tube. The bubbling frequency was adjusted to 0.1-1 Hz and we made sure there were at most a few bubbles at the interface without forming a bubble column. Each experiment ran for 48 hours to produce enough submicrometre droplets to be detected. To reduce the influence of dust, the container was made clean before each experiment. During the experiment, we reduced the contamination of the interface and the bulk by covering the tank. We collected samples deep in the bath and far from the interface. Samples were analyzed with DLS 8 hours after sampling without any further treatment. The high-speed camera was applied to capture the bubble-bursting process while ellipsometry was utilized for observation of the oil layer at the interface. The UV-cured experiments were carried out using a UV oven (IntelliRay 400, Uvitron) to crosslink the particles. The UV wavelengths were within the range 320–390 nm and the exposure time was 15 s.

**Materials** An aqueous surfactant solution was used as the continuous phase. Ultrapure water (resistivity 18.2 MΩ, Millipore MilliQ) was used to prepare all solutions. We used the surfactants $C_{16}$TAB (Hexadecyltrimethylammonium bromide,

Sigma-Aldrich, BioXtra, ≥99%), $C_{12}TAB$ (Dodecyltrimethylammonium bromide, Sigma-Aldrich, BioXtra, ~99%) and Aerosol OT (Docusate sodium, Sigma-Aldrich) were applied in the experiments as the water-soluble surfactants. For the oil phase, we used n-hexadecane (Sigma-Aldrich, anhydrous, ≥ 99%, $\rho$ = 773 kg/m$^3$, $\eta$ = 3.03 mPa·s at 25°C), n-pentadecane (Sigma-Aldrich, ≥ 99%, $\rho$ = 769 kg/m$^3$, $\eta$ = 2.56 mPa·s at 25°C), n-tetradecane (Sgima-Aldrich, olefine free, ≥ 99%, $\rho$ = 762 kg/m$^3$, $\eta$ = 2.10 mPa·s at 25°C), n-tridecane (Sigma-Aldrich, ≥ 99%, $\rho$ = 756 kg/m$^3$, $\eta$ = 1.71 mPa·s at 25°C), n-dodecane (Sigma-Aldrich, anhydrous, ≥ 99%, $\rho$ = 750 kg/m$^3$, $\eta$ = 1.38 mPa·s at 25°C), n-undecane (Sigma-Aldrich, ≥ 99%, $\rho$ = 740 kg/m$^3$, $\eta$ = 1.15 mPa·s at 25°C), n-decane (Sigma-Aldrich, anhydrous, ≥ 99%, $\rho$ = 730 kg/m$^3$, $\eta$ = 0.92 mPa·s at 25°C), and poly(dimethylsiloxane) (Sigma-Aldrich, $\upsilon$ = 1, 5 or 10 cSt at 25°C). The UV-cured material in Figs. 4d and e is Norland Optical Adhesive 89 which is cured by ultraviolet light with maximum absorption within the range of 310-395 nm.

**High-speed imaging** A high-speed camera (Vision Research, Phantom V7.3) with a lens (Sigma, DG Macro 105 mm) was used to record high-speed videos of the bubble collapse, at frame rates from 6800 up to 32000 fps. Movies were processed using Fiji software.

**Dynamic light scattering** The size of the submicrometre droplets was determined by dynamic light scattering (DLS) using a Malvern Zetasizer Nano ZS. The measurements were performed at 12.8° or 173° scattering angle with 4 mW He-Ne laser producing light with wavelength of 633 nm. DLS data were processed with Malvern's software using a distribution analysis based on a cumulant model to fit a single exponential to the correlation function to obtain the cumulant mean size and size distribution of the submicrometre droplets. The cumulant analysis is defined in ISO standard document 13321. The calculations of PDI are defined in the ISO standard document 13321:1996 E. Results of the PDI in different measurements were shown in Supplementary Materials.

**Ellipsometry** Ellipsometry experiments are carried out in the following way: polarized laser light (with wavelength 532 nm) was shined at the surface of a Petri dish with an aqueous solution of $[C_{16}TAB]$ = 0.9 mM and the reflected signal was recorded with a detector. The setup is based on a null type ellipsometer (LEF 3M, Novosibirsk, Russia), equipped with a rotating analyzer unit that allows to measure the changes in reflected light polarization in time steps of ~0.2 seconds. The angle of incidence is 50.0°. The instrument records the ellipsometric angles $\Psi$, $\Delta$, where $\tan\Psi \exp(i\Delta)$ is the polarization ratio of the output to the input signal. The technique is described in detail elsewere[31]. As the laser spot has a finite dimension (~1 mm$^2$), the values of $\Psi$, $\Delta$ are the average ones for the entities present in this spot. When stable base lines at the air/aqueous $C_{16}TAB$ boundary were established, a drop of 10 μL hexadecane was carefully added on the interface (far from the laser beam), and changes in polarization were detected, which in turn allow the calculation of the film thickness (assuming a refractive index of hexadecane n = 1.4340).

320

## Acknowledgments

322    We acknowledge the contribution of S. C. Russev from Department of Solid State Physics & Microelectronics, University of Sofia, Bulgaria who helped us with the interpretation of the ellipsometric data and R. D. Stanimirova from Department of Chemical Engineering, University of Sofia, Bulgaria, who performed measurements in a Langmuir trough and some spreading experiments. T. D. G. and S. D. S. acknowledge the financial support of EU project FP7-REGPOT-2011-1, "Beyond Everest". M. R. acknowledges D. Langevin for fruitful discussions. H. A. S. thanks the NSF for support via grant CBET.

330

331

332

**Figure Legends**

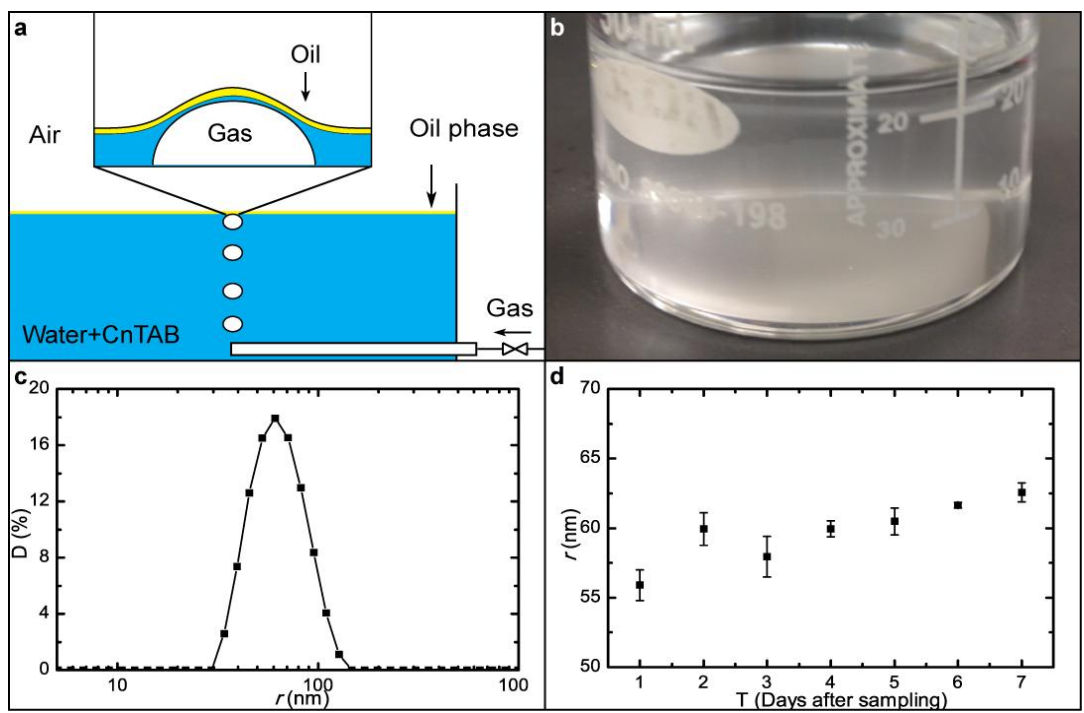

**Figure 1 | Bubble bursting at an air/oil/water interface.** (a) Sketch of the experimental system. Inset: close-up of the deformed compound interface. (b) Image of the translucent aqueous phase after bubbling for 48 hours (Oil phase: dodecane, aqueous phase: $[C_{16}TAB]$ = 0.09 mM). (c) Size distribution of the oil droplets based on the intensity measured by DLS (Oil phase: hexadecane, initial thickness of the oil layer $h_I$ = 1 mm; bubble diameter $d_b$ = 2.8 ± 0.25 mm, aqueous phase: $[C_{16}TAB]$ = 0.09 mM). The peak value of the distribution (59.8 nm here, with PDI = 0.091) is taken as the radius of the submicrometre droplets. (d) Time evolution of the size of the submicrometre droplets in the same sample shown in Fig. 1c over a week, which demonstrates stability of the submicrometre droplets. The error bar here is defined as the standard deviation of the droplet size in three DLS measurements.

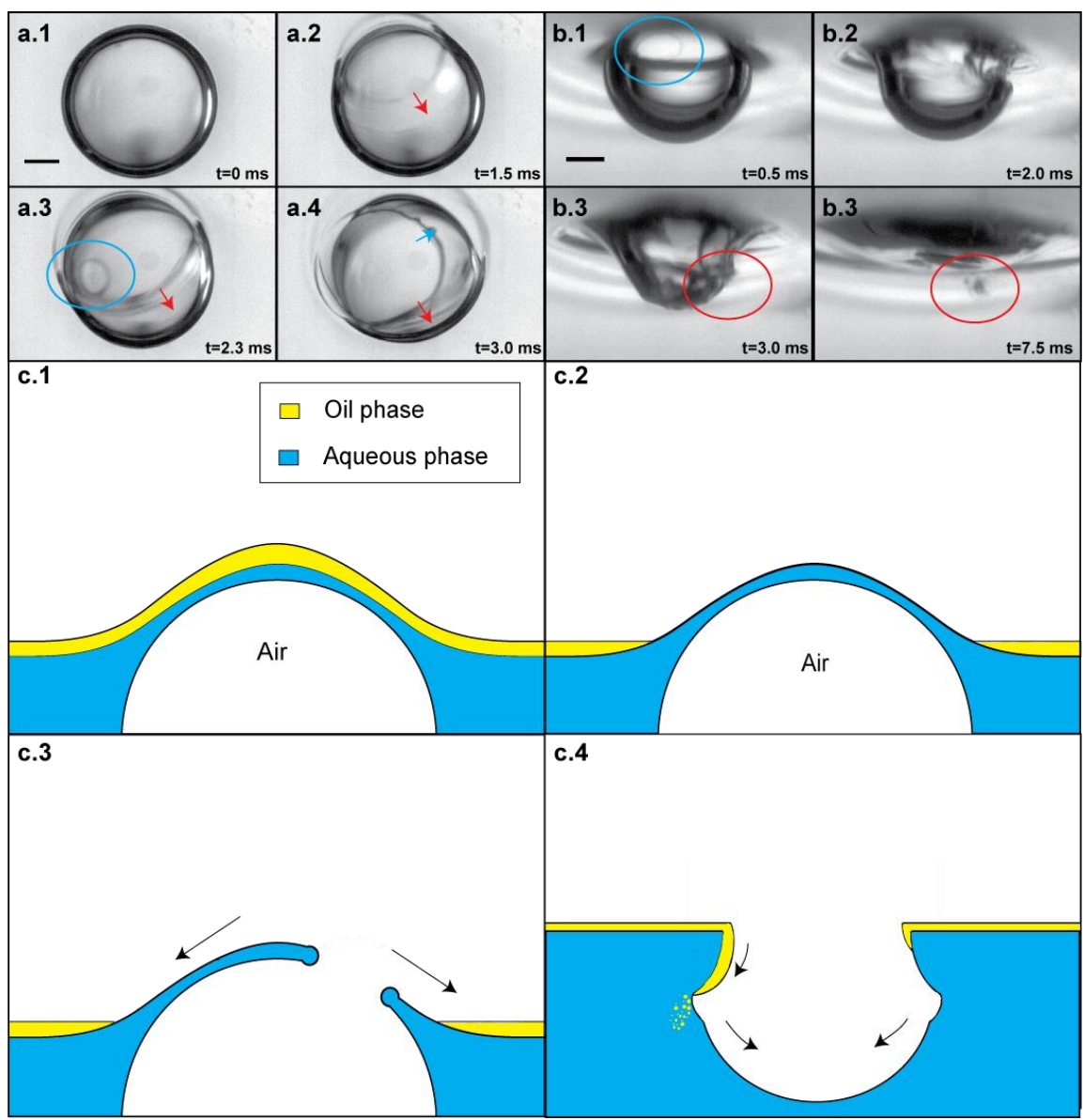

**Figure 2 | High-speed observations of the bursting process and schematic descriptions for the dispersal mechanism.** (a) Top-view photos of the bursting of a bubble at interfaces of air/hexadecane/water at *[$C_{16}TAB$]* = 0.09 mM ($h_I$ = 1 mm; $d_b$ = 4.0 ± 0.21 mm; scale bar is 1 mm): a.1, The bubble rests near the compound interface with oil and water films on top of the bubble; a.2, The oil film (above the water film) ruptures first before the water film. The retraction direction of the oil film is shown with the red arrow; a.3, After the rupture of the oil film, a hole opens in the water film, as shown with the blue circle; a.4, The water film then retracts along a direction different from that of the oil film. The retraction direction of the water film is shown with the blue arrow. b. Side-view photos of the bursting of a bubble at interfaces of air/hexadecane/water at *[$C_{16}TAB$]* = 0.09 mM ($h_I$ = 1 mm; $d_b$ = 4.0 ± 0.21 mm; scale bar is 1 mm). b.1, A hole is nucleated on the cap of the bubble, as shown with the blue circle; b.2, The surface of the cavity deforms; b.3, The deformation propagates further down the interface; b.4, A

spray of droplets is created at the wall of the cavity, as shown with the red circle. Note that (a) and (b) are not taken simultaneously. c. Sketch of mechanism for the dispersal formation.

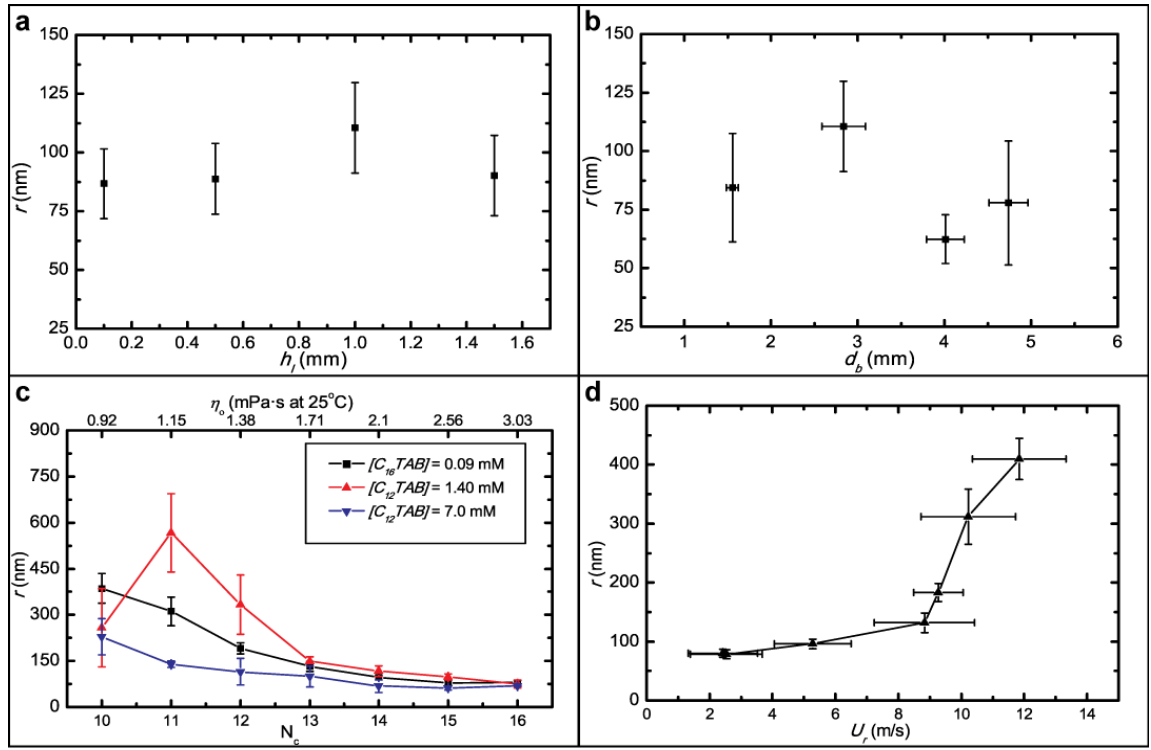

**Figure 3 | Influence of oil thickness ($h_I$), bubble diameter ($d_b$), viscosity of oil ($\eta_o$) and carbon number ($N_c$) of the oil on the size of the submicrometre-sized droplets ($r$).** (a) Relationship between $r$ and $h_I$ (Oil phase: hexadecane; $d_b = 2.8 \pm 0.25$ mm; aqueous phase: $[C_{16}TAB] = 0.09$ mM). (b) Relationship between $r$ and $d_b$ (Oil phase: hexadecane; $h_I = 1$ mm; aqueous phase: $[C_{16}TAB] = 0.09$ mM). (c) Relationship between $r$ and $\eta_o$ as well as $N_c$ (Oil phase: hexadecane; $h_I = 1$ mm; $d_b = 2.8 \pm 0.25$ mm; aqueous phase: $[C_{16}TAB] = 0.09$ mM and $[C_{12}TAB] = 1.40$ mM or $7.0$ mM. Note that $r$ for $N_c = 11$ with $[C_{12}TAB] = 1.40$ mM was determined by microscope using image analysis since DLS could not obtain a reliable correlation function for these samples. (d) Relationship between $r$ and the speed of the last receding film $U_r$.

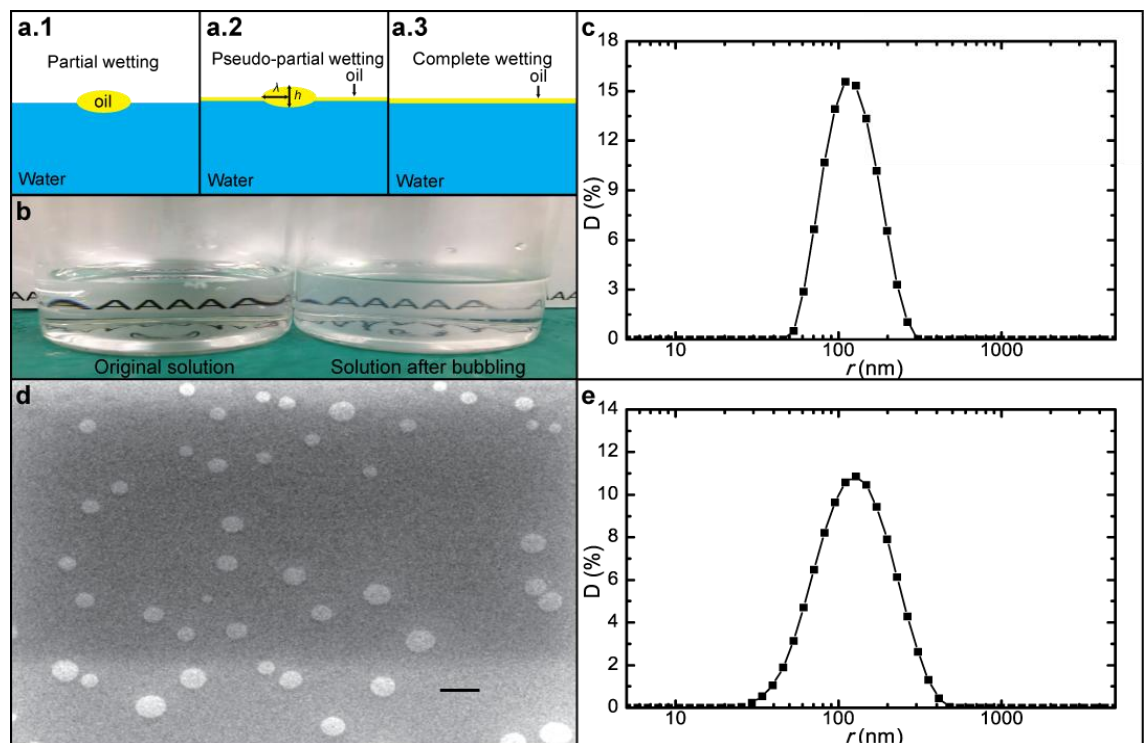

**Figure 4 | Sketch of different wetting states and formation of petroleum dispersal and polymeric submicrometre particles.** (a) Partial wetting, pseudo-partial wetting and complete wetting behaviors of an oil drop on an aqueous surface. The partial wetting state is characterized by an oil lens sitting at the air/water interface in equilibrium with a two-dimensional dilute gas of oil molecules while, in the pseudo-partial wetting state, the oil initially spreads after deposition and then forms lenses in equilibrium with a microscopic film of a few molecules thick. The oil spreads out to form a film of uniform thickness covering the whole surface in the complete wetting state. (b) Image of the aqueous phase after bubbling for 48 hours using petroleum as the oil phase ($h_I = 1$ mm; $d_b = 2.8 \pm 0.25$ mm; aqueous phase: $[C_{16}TAB] = 0.09$ mM). The solution after bubbling is hazy compared with the original solution. (c) Results of DLS measurement for samples of the solution after bubbling. The size of the droplets is 113.4 nm with PDI = 0.101. (d) ESEM photo of UV-cured submicrometre particles produced by bubble bursting (Oil phase: Norland Optical Adhesive (NOA) 89; $h_I = 1$ mm; $d_b = 2.8 \pm 0.25$ mm; aqueous phase: $[C_{16}TAB] = 0.09$ mM). Scale bar is 500 nm. (e) DLS results of the NOA 89 sample before the UV-cured process. The size of the particles is 109.8 nm with PDI = 0.197.
21